\documentclass[a4paper]{article}
\usepackage{graphicx}
\usepackage{amsmath}

\begin{document}

\begin{center}
{\bf \Large Node-node distance distribution for growing networks\footnote{presented at the 37-th Polish Physicists' Meeting, Gda\'nsk, Poland, 15-19 Sep. 2003}}\\[5mm]

{\large K.Malarz$^*$, J.Karpi\'nska, A.Kardas, and K.Ku{\l}akowski$^\dag$}

\bigskip

{\em Faculty of Physics and Nuclear Techniques\\
AGH University of Science and Technology\\
al. Mickiewicza 30, PL-30059 Krak\'ow, Poland\\

\bigskip

E-mail:
$^*${\tt malarz@agh.edu.pl}, $^\dag${\tt kulakowski@novell.ftj.agh.edu.pl}

\bigskip

\today
}
\end{center}

\begin{abstract}
\noindent
We present the simulation of the time evolution of the distance matrix.
The result is the node-node distance distribution for various kinds of networks.
For the exponential trees, analytical formulas are derived for the moments of the distance distribution.
\end{abstract}

\section{Introduction}

	A graph is defined as a set of nodes (vertices) and a set of links between nodes (edges) \cite{clark,wilson,berge,harary}.
By graph evolution or growth we mean subsequent attaching of new nodes with $m$ edges to previously existing nodes \cite{dorogovtsev3}.
Such growing graphs may reflect some features of real evolving networks, e.g. a network of collaborators, a network of citations of scientific papers, some biological networks (food chains or sexual relations) or Internet and world-wide-web pages with links between them \cite{dorogovtsev3,dorogovtsev1,dorogovtsev2,newman,albert}.

	The distance between nodes is the shortest number of edges which leads from one node to the other.
The node-node distance (NND) distribution depends on how subsequent nodes are attached.
If each node is connected with {\em only one} of preexisting nodes ($m=1$) {\em a tree} appears.
When a new node is attached to several different nodes with $m>1$ edges, the growing structure is called {\em a simple graph}.
We may choose nodes to which new nodes are attached in preferential way or randomly.
In the latter case we deal with {\em exponential} networks. 
If the probability of choosing a node is proportional to its degree (e.g. to the number of its nearest neighbors) the growing structure is called {\em scale-free} or Albert--Barab\'asi networks \cite{barabasi}.

	In this paper, the numerical algorithm for the network growth --- basing on {\em distance matrix} evolution --- is presented both, for exponential and scale-free networks ($m=1, 2$) \cite{malarz1,malarz2}.
The NND distribution and its characteristics are calculated.
For the exponential trees the iterative formulas for $n$-th ordinary moments are derived.

\section{Computer simulations}

        A graph with edges of unit length may be fully characterized by its 
distance matrix $\mathbf{S}$, an element $s_{ij}$ of which is equal to the 
shortest path between nodes $i$ and $j$.
This matrix representation is also particularly useful when computer simulations for graph evolution are applied.

        Attaching subsequent node with one edge ($m=1$) to previously 
existing network of $N$ nodes corresponds to adding a new $(N+1)$-th 
row and a new column to $N\times N$ large distance matrix $\mathbf{S}$.
The distance from newly added $(N+1)$-th node to all others via selected ---
labeled as $p$ --- node is larger by one than the distance between $p$-th and all others.
Thus, new $(N+1)$-th row/column is a simple copy of $p$-th row/column but 
with all of its elements incremented \cite{malarz1}:
\begin{equation}
\label{eq_trees}
\forall ~ 1\le i \le N: s_{N+1}(N+1,i)=s_{N+1}(i,N+1)=s_N(p,i)+1.
\end{equation}

        Similarly, when new node is attached to the network with two edges 
($m=2$) to two different nodes --- which are labeled as $p$ and $q$ --- but the 
distance from all other nodes $i$ to the newly added $(N+1)$-th is one plus 
the smaller distance between $p$--$i$ or $q$--$i$ nodes pairs \cite{malarz2}:
\begin{equation}
\label{eq_graphs_1}
\forall ~ 1\le i\le N: s_{N+1}(N+1,i)=s_{N+1}(i,N+1)=\min\big(s_N(p,i),s_N(q,i)\big)+1.
\end{equation}

        In the case mentioned above of the growth of the simple graphs also the 
reevaluation of distances between nodes $i$ and $j$ must be done to check if 
adding a new node provides the shortcut \cite{malarz2}:
\begin{equation}
\begin{split}
\label{eq_graphs_2}
\forall ~ 1\le i,j\le N: s_{N+1}(i,j)=\\
\min\big(s_N(i,j),s_N(i,p)+2+s_N(q,j),s_N(i,q)+2+s_N(p,j)\big).
\end{split}
\end{equation}

        In both cases diagonal elements of new row/column are zero \cite{malarz1,malarz2}:
\begin{equation}
\label{eq_diagonal}
s_{N+1}(N+1,N+1)=0.
\end{equation}

        Selecting rows/columns (nodes to which we attempt to add a new node) 
may be random or preferential.
In the latter case an additional evolving vector is introduced, which contains the node labels.
These labels occur as vector elements with a probability proportional to the degree of the node.
Random selection of elements of such a vector correspond to Albert--Barab\'asi 
construction rule.
The procedure is known as the Kert\'esz algorithm \cite{kertesz}.

\section{Analytical calculations}

	Let us define $n$-th moments of the NND distribution for all distances
\begin{equation}
\ell_N^n\equiv [\{ s^n(i,j) \}] =\dfrac{1}{N^2}\sum_{i=1}^N\sum_{j=1}^N [s^n(i,j)],
\label{eq_ave}
\end{equation}
and only for non-zero distances
\begin{equation}
d_N^n\equiv [\langle s^n(i,j) \rangle] =\dfrac{1}{N(N-1)}\sum_{i=1}^N\sum_{\substack{j=1\\j\ne i}}^N [s^n(i,j)],
\label{eq_mom}
\end{equation}
where $\{\cdots\}$, $\langle\cdots\rangle$ and $[ \cdots ]$ denote an average 
over $N^2$ matrix elements, an average over $N(N-1)$ non-diagonal matrix elements, and an average over $N_{\text{run}}$ independent realizations of the evolution process (matrices), respectively.
Moments \eqref{eq_ave} and \eqref{eq_mom} for $n=1$ are sometimes called {\em the network diameter}. 
Both double sums in r.h.s. of Eqs. \eqref{eq_ave} and \eqref{eq_mom} are equal, due to obvious fact, that $s(i,i)=0$. 
That allows to derive simple dependence between averages $\{\cdots\}$ 
and $\langle\cdots\rangle$:
\begin{equation}
N \ell_N^n = (N-1) d^n_N.
\label{eq_ell_d}
\end{equation}
For the exponential trees --- basing on $s(i,i)=0$ and distance matrix symmetry $s(i,j)=s(j,i)$ --- we are able to construct iterative equations for $\ell^n_{N+1}$ as dependent on $\ell^k_N$ ($k=1,\cdots,n$)
\begin{equation}
(N+1)^2 \ell^n_{N+1}=\sum_{i=1}^{N+1} \sum_{j=1}^{N+1} [s^n(i,j)]=
N^2\ell^n_N+2\sum_{i=1}^N \big(1+[s(i,q)]\big)^n,
\label{eq_main}
\end{equation}
where $q$ is the number of the randomly selected row/column of the distance matrix $\mathbf{S}$.
Combining Eq. \eqref{eq_main} with Eq. \eqref{eq_ell_d} gives the desired iterative formula:
\begin{equation}
d_{N+1}^n=\dfrac{(N+2)(N-1)}{(N+1)N}d_N^n+\dfrac{2}{N+1}+\dfrac{2(N-1)}{(N+1)N}
\sum_{k=1}^{n-1} {n \choose k}d_N^k.
\label{eq_itr}
\end{equation}

\section{Results and conclusions}

	For the trees the mean of the NND $d_N^1$ and its dispersion 
$\sigma^2\equiv d_N^2-(d_N^1)^2$ grow logarithmically with $N$ (see Tabs. \ref{tab1}, \ref{tab2}) \cite{malarz1}. 
For the graphs only the first cumulant (the average of the NND $d_N^1$) 
grows logarithmically (see Tab. \ref{tab2}) \cite{malarz2}.
Such a slow increase of $d_N^1$ with number of network nodes is known as the small-world effect \cite{milgram}.
\begin{center}
\begin{table}
\begin{center}
\caption{The mean distance $d(N)=a\ln N+b$ for different evolving networks.}
\label{tab1}
\begin{tabular}{r cc cc} 
\hline
    & exponential & exponential & scale-free & scale-free \\ 
\hline
$m$ & 1       & 2     & 1       & 2     \\
$a$ & 2.00    & 0.672 & 1.00    & 0.462 \\
$b$ & $-2.84$ & 0.296 & $-0.08$ & 0.889 \\ 
\hline
\end{tabular}
\end{center}
\end{table}
\end{center}
\begin{center}
\begin{table}
\begin{center}
\caption{The dispersion $\sigma^2(N)=c\ln N+d$ for the exponential and the scale-free trees ($m=1$).}
\label{tab2}
\begin{tabular}{r cc} 
\hline
& exponential & scale-free \\ 
\hline
$c$ & 2.00    & 1.00       \\
$d$ & $-1.44$ & $-1.64$    \\ 
\hline
\end{tabular}
\end{center}
\end{table}
\end{center}
\begin{figure}
\begin{center}
\includegraphics[angle=-90,width=.9\textwidth]{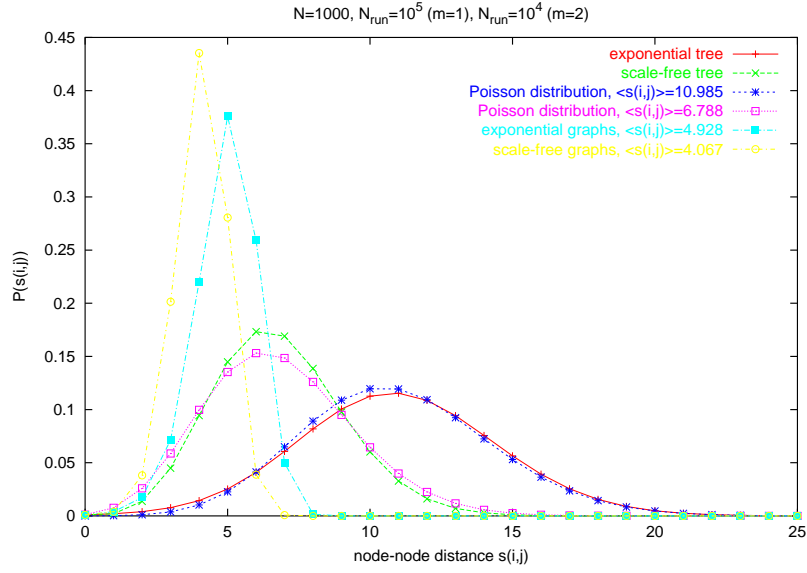}
\end{center}
\caption{The NND distribution for different types trees and graphs.}
\label{fig_his}
\end{figure}
\begin{figure}
\begin{center}
\includegraphics[angle=-90,width=.9\textwidth]{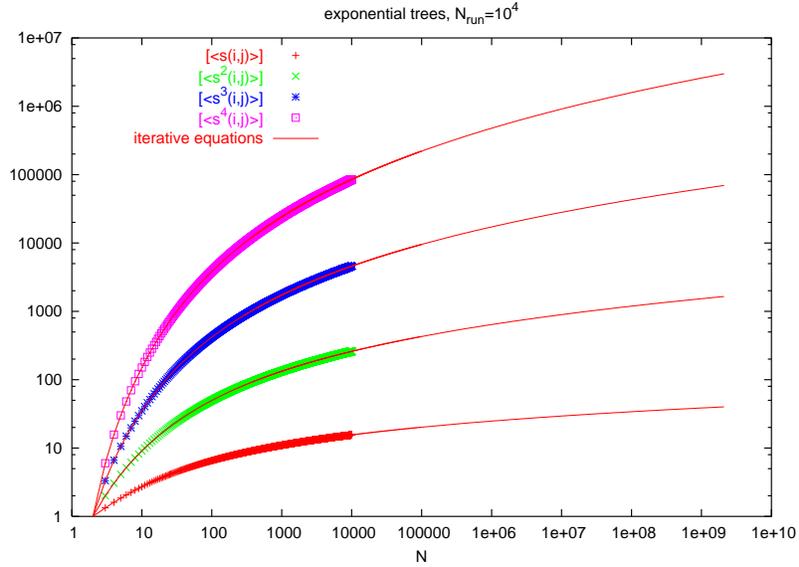}
\end{center}
\caption{The main moments $d^k_N$ ($k=1,\cdots,4$) for the exponential trees given by Eq. \eqref{eq_itr} (lines) and from the direct simulations (symbols).
The latter are averaged over $N_{\text{run}}=10^4$ independent evolution process realizations.}
\label{fig_mom}
\end{figure}
\begin{figure}
\begin{center}
\includegraphics[angle=-90,width=.9\textwidth]{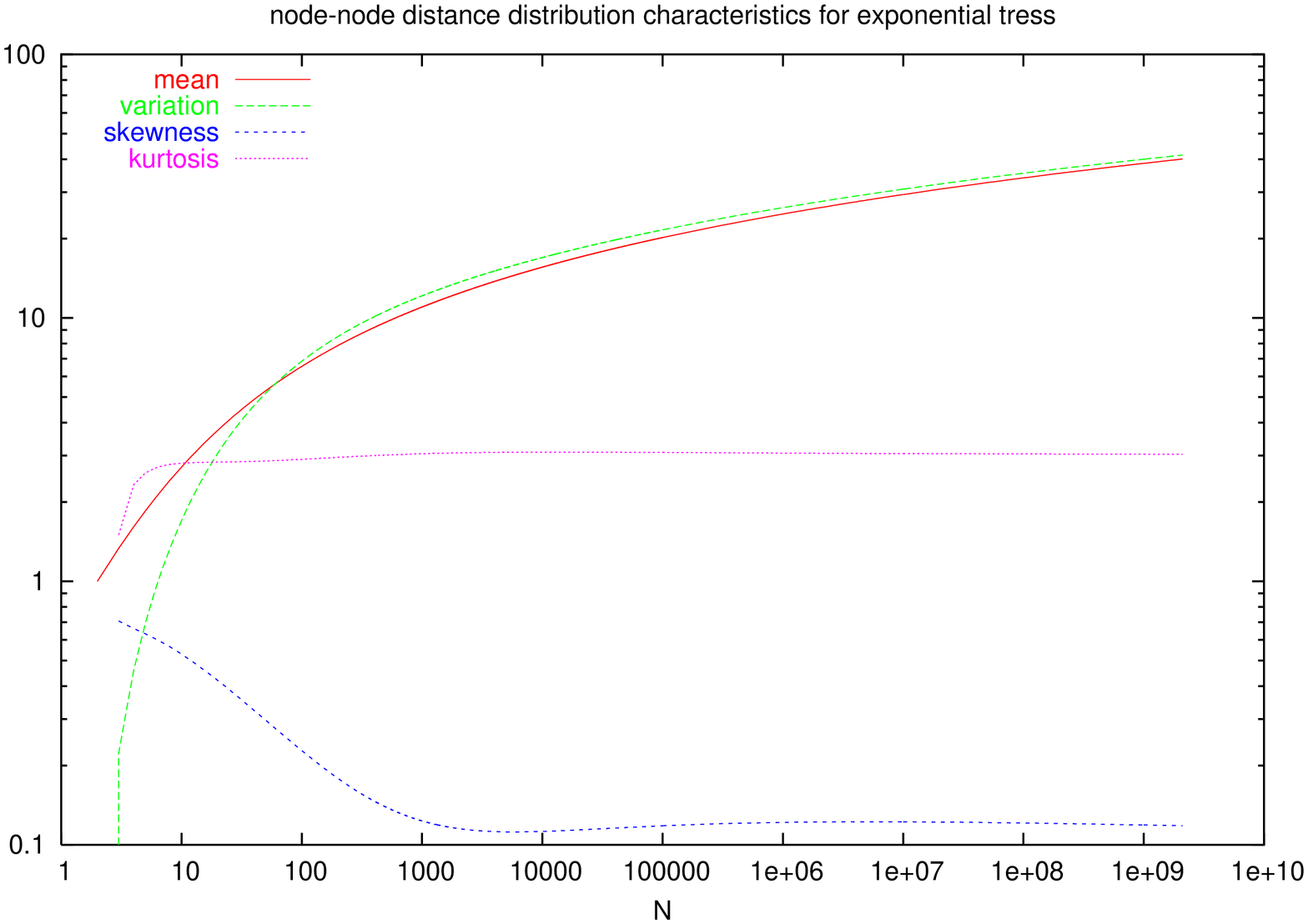}
\end{center}
\caption{The NND distribution characteristics for the exponential trees as derived from iterative Eq. \eqref{eq_itr}.}
\label{fig_sta}
\end{figure}

	The histogram of NND is presented in Fig. \ref{fig_his}.
	As we expected the NND for the graphs are more condensed than the NND for the trees, as well as, the scale-free graphs (trees) are more condensed than  the exponential graphs (trees).

	Knowing the moments $d^n_N$ --- the averages of the $n$-th powers of the non-diagonal distance matrix elements \eqref{eq_mom} --- allows to build all statistical parameters which characterize the NND distribution, e.g. the average distance $d$, the distance dispersion $\sigma^2$, its skewness
\[
\nu_3\equiv\dfrac{d_N^3-3d_N^2d_N^1+2(d_N^1)^3}{\sigma^3},
\]
and kurtosis 
\[
\kappa_4\equiv\dfrac{d_N^4-4d_N^3d_N^1+6d_N^2(d_N^1)^2-3(d_N^1)^4}{\sigma^4}.
\]
The values of such characteristics of the NND for the exponential trees obtained via Eq. \eqref{eq_itr} are presented in Figs. \ref{fig_mom} and \ref{fig_sta}.
For trees, the distributions are similar to the Poisson distribution (see Fig. \ref{fig_his}).
However, even for large $N$ the skewness and kurtosis do not vanish as one may expect for the normal distribution \cite{szabo}. 

\section*{Acknowledgments}
K.M.'s participation in the 37-th Polish Physicists' Meeting in Gda\'nsk was 
financed by Krak\'ow Branch of the Polish Physical Society (PTF).
The numerical calculations were carried out in ACK--\-CYF\-RO\-NET--\-AGH.
The machine time on SGI~2800 is financed by the State Committee for 
Scientific Research (KBN) in Poland with grant
No. KBN/\-SGI2800/\-AGH/\-018/\-2003.



\end{document}